\documentclass[
prc,%
10pt,%
final,%
notitlepage,%
oneside,%
twocolumn,%
nobibnotes,%
nofootinbib,
superscriptaddress,%
floatfix,%
showkeys,%
showpacs]%
{revtex4}
\usepackage{color}
\usepackage{amsfonts}
\usepackage{amsbsy}
\usepackage{mathrsfs}
\usepackage{graphicx}
\def\lsim{\mathrel{\rlap{
\lower4pt\hbox{\hskip-3pt$\sim$}}
    \raise1pt\hbox{$<$}}}     
\def\gsim{\mathrel{\rlap{
\lower4pt\hbox{\hskip-3pt$\sim$}}
    \raise1pt\hbox{$>$}}}     
\def\scr#1{\mbox{\scriptsize #1}}
\begin{document}
\title{
Elliptic Flow in Heavy-Ion Collisions at Energies $\sqrt{s_{NN}}=$ 2.7--39 GeV} 
\author{Yu. B. Ivanov}\thanks{e-mail: Y.Ivanov@gsi.de}
\affiliation{National Research Centre "Kurchatov Institute" (NRC "Kurchatov Institute"), 
Moscow 123182, Russia}
\affiliation{National Research Nuclear University "MEPhI"  (Moscow Engineering
Physics Institute),
Moscow 115409, Russia}
\author{A. A. Soldatov}\thanks{e-mail: saa@ru.net}
\affiliation{National Research Nuclear University "MEPhI"  (Moscow Engineering
Physics Institute),
Moscow 115409, Russia}
%
%
\begin{abstract}
The 
transverse-momentum integrated elliptic flow of charged particles
at midrapidity, $v_2$(charged), and that of identified hadrons from Au+Au collisions  
are computed in a wide range of incident energies
 2.7 GeV  $\le \sqrt{s_{NN}}\le$ 39 GeV. The simulations are performed  
within a three-fluid model 
employing three different equations of state (EoS's): a purely hadronic EoS   
and two versions of the EoS involving the   deconfinement
 transition---a first-order phase transition  and a smooth crossover one. 
The present simulations  demonstrate low sensitivity of $v_2$(charged)
to  the EoS. All considered scenarios equally well reproduce 
recent STAR data on $v_2$(charged)
for mid-central Au+Au collisions and  properly describe its 
change of sign at the incident energy decrease below 
$\sqrt{s_{NN}}\approx$ 3.5 GeV. 
The predicted integrated elliptic flow of various species exhibits 
a stronger dependence on the EoS. 
A noticeable sensitivity to the EoS is found for anti-baryons 
and, to a lesser extent, for $K^-$ mesons. 
In particular, the $v_2$ excitation functions of  anti-baryons 
 exhibit a non-monotonicity within the deconfinement
scenarios that was predicted  by Kolb, Sollfrank and Heinz. 
However, low multiplicities of  anti-baryons 
at $\sqrt{s_{NN}}\leq$ 10 GeV  
result in large fluctuations of their $v_2$ which may wash out this non-monotonicity. 
\pacs{25.75.-q,  25.75.Nq,  24.10.Nz}
\keywords{relativistic heavy-ion collisions, elliptic flow,
  hydrodynamics, deconfinement}
\end{abstract}
\maketitle

\section{Introduction}

The Beam Energy Scan (BES) program at the Relativistic Heavy Ion Collider
(RHIC) at Brookhaven National Laboratory (BNL) pursues the major goal of exploring the QCD
phase diagram of the strongly interacting matter. 
The main questions addressed in this research are:  At which energy does onset of deconfinement happen? 
What is the order of the deconfinement transition at high baryon densities? 
Is there a critical end point in the phase diagram? 
The BES program at RHIC  provides us with a unique opportunity to study systematically the collision
energy dependence of a large number of observables.
The present study is inspired by recent papers of the STAR Collaboration 
\cite{Adamczyk:2012ku,Adamczyk:2013gw} on the beam-energy dependence of the 
elliptic flow ($v_2$) in the BES region.

The beam-energy dependence of the collective flow has been recently studied within several
different models 
\cite{Kolb:2000sd,Kestin:2008bh,Petersen:2009vx,Petersen:2010md,Auvinen:2013sba,Shen:2012vn,Solanki:2012ne,Plumari:2013bga,Konchakovski:2012yg,Ivanov:2013mxa,Steinheimer:2012bn}
with the main emphasis on search of signals of the onset of deconfinement. 
A non-monotonicity of the transverse-momentum integrated 
($p_t$-integrated) $v_2$ was predicted in Ref. \cite{Kolb:2000sd}
which is related to the quark-hadron phase transition and the corresponding
softening of the EoS in the transition region. However, later
it was stated \cite{Kestin:2008bh} that in the experimental data this phase
transition signature will be washed out by strong viscous effects in the late hadronic phase,  
where the fireball spends most of its time. As a result, the
experimentally measured integrated elliptic flow $v_2$ should rise monotonically with 
the center-of-mass energy, $\sqrt{s_{NN}}$, approaching
the ideal fluid limit only at or above RHIC energies.

On the other hand, in Ref. \cite{Denicol:2013nua} it was 
found that the hadron resonance gas with a large baryon density is closer to the ideal
fluid limit than the corresponding gas with zero baryon density. 
Moreover, a nonzero baryon chemical potential serves not
only to reduce the effect of dissipative terms of the first order in gradients 
but also of the second-order terms.
This effect of the baryon chemical potential was noticed even earlier in Ref. \cite{Khvorostukhin:2010aj}. 
The latter suggests that the system created at lower collision energies may display a fluid-like behavior with an
effective fluidity close to that found at RHIC top-energy collisions, thus explaining why the differential
elliptic flow measured at lower RHIC energies  is close to that observed  
at the top RHIC energies. 
Indeed, an effective fluidity  extracted from experimental data on collective flow \cite{Lacey:2013qua}
indicates that the viscosity at lower BES-RHIC energies is only slightly higher than that 
at the top RHIC energy. 
Above findings were supported by actual simulations within a hybrid model \cite{Auvinen:2013sba}. 
It was found that the triangular flow provides the clearer signal for the formation of low-viscous 
fluid in heavy ion collisions. 
Moreover, the kinetic phase produces additional elliptic flow rather than destroy it
which also testify in favor of low-viscous fluid at low BES energies.

A the same time in Ref. \cite{Petersen:2009vx} it was pointed out 
a strong influence of initial conditions for the hydrodynamic evolution on
the observed $v_2$ values, thus questioning the standard interpretation that the hydrodynamic
limit is only reached at RHIC energies.
The integrated and differential elliptic flow for charged particles at SPS
energies was found to be mostly sensitive to viscosity rather than to
the EoS \cite{Petersen:2010md}.
Recently, a low sensitivity of the elliptic flow to the type of the phase transition (or its absence) 
in the EoS was already reported \cite{Dudek:2014qoa}
at RHIC energies.  
Thus, the situation is somewhat controversial and needs further investigation.

In the present paper we report results on the collision energy dependence of 
the 
midrapidity transverse-momentum integrated elliptic flow 
of charged particles 
and that of identified hadrons
produced in Au+Au collisions   
using a model of the three-fluid 
dynamics (3FD) \cite{3FD} employing three different equations of state (EoS): 
a purely hadronic EoS
\cite{gasEOS} (hadr. EoS)
that was used in calculations of the collective flow so far 
\cite{3FDflow,3FD-GSI07,3FDv2}
and two versions of EoS involving the deconfinement 
transition \cite{Toneev06}. These two versions are an EoS with the first-order phase transition 
(2-phase EoS)
and that with a smooth crossover transition
(crossover EoS). 
We report results of simulations 
in the energy range from 2.7 GeV 
to 39 GeV in terms of  $\sqrt{s_{NN}}$. This domain 
goes beyond the range of the RHIC BES program and also covers energies of the Alternating Gradient
Synchrotron (AGS) at BNL 
and the  
Super Proton Synchrotron (SPS)
of the European Organization for Nuclear Research (CERN).  
The reported results are also relevant to newly constructed 
Facility for Antiproton and Ion Research (FAIR) in Darmstadt and the
Nuclotron-based Ion Collider Facility (NICA) in Dubna.

The 3FD model does not include viscosity in its formulation.
However, dissipation is present in the 3FD trough friction interaction 
between participated fluids. 
It would be interesting to estimate this dissipation in terms of viscosity
of the conventional 1-fluid hydrodynamics. This work is in progress. 
Numerical solution of the model
is performed in (3+1) dimensions on spacial grid  
with a finite cell size. The numerical scheme (the particle-in-cell method)
has a relatively small numerical viscosity,   
which is comparable to the minimal viscosity that occurs at 
deconfinement transition \cite{Csernai:2011qq}\footnote{
We use the same numerical scheme as that in Ref. \cite{Csernai:2011qq}. 
}.
Details of these calculations are described in Ref.  
\cite{Ivanov:2013wha} dedicated to analysis of the baryon stopping. 

We would like to mention explicitly that no tuning (or
change) of 3FD-model parameters
or parameters of the used EoS's
 has been performed in this study as compared to 
previous simulations 
\cite{Ivanov:2013mxa,Ivanov:2013wha,Ivanov:2012bh,Ivanov:2013yqa,Ivanov:2013cba,Ivanov:2013yla}
in which various bulk observables were considered  
precisely in the same range of incident energies. The main goal of this series of works
\cite{Ivanov:2013mxa,Ivanov:2013wha,Ivanov:2012bh,Ivanov:2013yqa,Ivanov:2013cba,Ivanov:2013yla}
is to analyze a whole set of various 
observables within the same description without any 
observable-dependent tunning of the parameters. 
It enables us to reveal possible correlations 
in excitation functions of various observables within
different scenarios, which can be used as experimental
indications of the deconfinement onset or its absence.

\section{The 3FD Model}
\label{Comparison}

A three-fluid approximation is a minimal way to 
simulate a finite stopping power of colliding nuclei at high incident energies.
Within this approximation 
a generally nonequilibrium distribution of baryon-rich
matter is modeled by counter-streaming baryon-rich fluids 
initially associated with constituent nucleons of the projectile
(p) and target (t) nuclei. In addition, newly produced particles,
populating the midrapidity region, are associated with 
a separate net-baryon-free fluid which is called a
``fireball'' fluid (f-fluid), following the Frankfurt group \cite{Kat93,Brac97}.
A certain formation time $\tau$ is allowed for the f-fluid, during
which the matter of the fluid propagates without interactions. 
The formation time 
is  associated with a finite time of string formation. It is similarly 
incorporated in kinetic transport models such as 
the Ultra-relativistic
Quantum Molecular Dynamics (UrQMD) \cite{Bass98},  
the Hadron string Dynamics (HSD)
\cite{Cassing99} and the Parton-Hadron String Dynamics (PHSD) \cite{CB09}. 
Each of these fluids (the f-fluid after its formation) 
is governed by conventional hydrodynamic equations
which contain interaction terms in their right-hand sides. 
These interaction terms describe mutual friction of the fluids and 
production of the f-fluid.
The friction between fluids was fitted to reproduce
the stopping power observed in proton rapidity distributions for each EoS, 
as it is described in  Ref. \cite{Ivanov:2013wha} in detail.
The main difference concerning the 
f-fluid in considered alternative scenarios
consists in different formation times: $\tau$ = 2 fm/c for the
hadronic scenario and
$\tau$  = 0.17 fm/c for scenarios involving 
the deconfinement transition \cite{Ivanov:2013wha}.
Large formation time within the hadronic scenario was chosen in order 
to reproduce mesonic yields at SPS energies.
This was done in line with a principle of fair treatment of any EoS: 
any possible uncertainties in the parameters are treated in favor of the EoS,
i.e. for each EoS the dynamical parameters of the model are chosen (within their uncertainty range) 
in such a way that  
the best possible reproduction of observables is achieved with this EoS.

Since each fluid is governed by its own hydrodynamic equations, 
it is locally characterized by its own set of hydrodynamic quantities---baryon and 
energy densities, velocities and, in particular, temperature and baryon chemical 
potential---which, in general, differ from those of other fluids. 
At the same time, all three fluids are described by the same EoS (chosen for the 
simulation), of course, with the pressure and energy density taken at  
their specific values of the temperature and baryon chemical 
potential. At the initial stage of the reaction all three fluids coexist in the same 
space-time region, thus describing a certain {\em nonequilibrium} state 
of the matter. It may happen that one or two of the fluids occur in the 
quark-gluon phase while other(s) is(are) in the hadronic one. This is a kind 
of a nonequilibrium mixed phase that is also possible in the model. 

At the final stage of the collision the p- and t-fluids are either spatially separated or unified, 
while the f-fluid still overlaps 
with the baryon-rich (p- and t-) fluids to a lesser (at high energies) 
or grater (at lower energies) extent. 
The freeze-out  is performed accordingly to the procedure described in Ref. \cite{3FD} 
and in more detail in Refs. \cite{Russkikh:2006aa,Ivanov:2008zi}. 
The freeze-out   criterion is based on a local    
energy density, $\varepsilon_{\scr{tot}}$,  defined as a sum of contributions 
from all (p-, t- and f-) fluids being present in a local space-time region. 
The freeze-out procedure starts if $\varepsilon_{\scr tot}<\varepsilon_{\scr{frz}}$.
The freeze-out energy density $\varepsilon_{\scr{frz}}$ = 0.4 GeV/fm$^3$ was chosen mostly
on the condition of the best reproduction of secondary
particle yields.
To the moment of the freeze-out the matter is already in the
hadronic phase in the case of the 2-phase EoS, while for
the crossover EoS this is not so. However, this is not
a problem because, any case, the thermodynamic quantities of the frozen-out matter are recalculated from the
in-matter EoS, with which the hydrodynamic calculation
runs, to the hadronic gas EoS\footnote{
In this gas EoS, as well as in the hadr. EoS, 48 different hadronic species are taken into
account. Each hadronic species includes all the relevant
isospin states, e.g., the nucleon species includes proton
and neutron.}. 
This is done because a part
of the energy is still accumulated in collective mean fields
at the freeze-out instant. This mean-field energy should
be released before calculating observables. Otherwise,
the energy conservation would be violated.
The freeze-out is performed locally (in a local spatial cell meeting 
the freeze-out condition), simultaneously for all fluids populating 
this cell. Thus, the freeze-out configuration of the matter is 
generally nonequilibrium.

%
\begin{figure}[tbh]
\includegraphics[width=5.8cm]{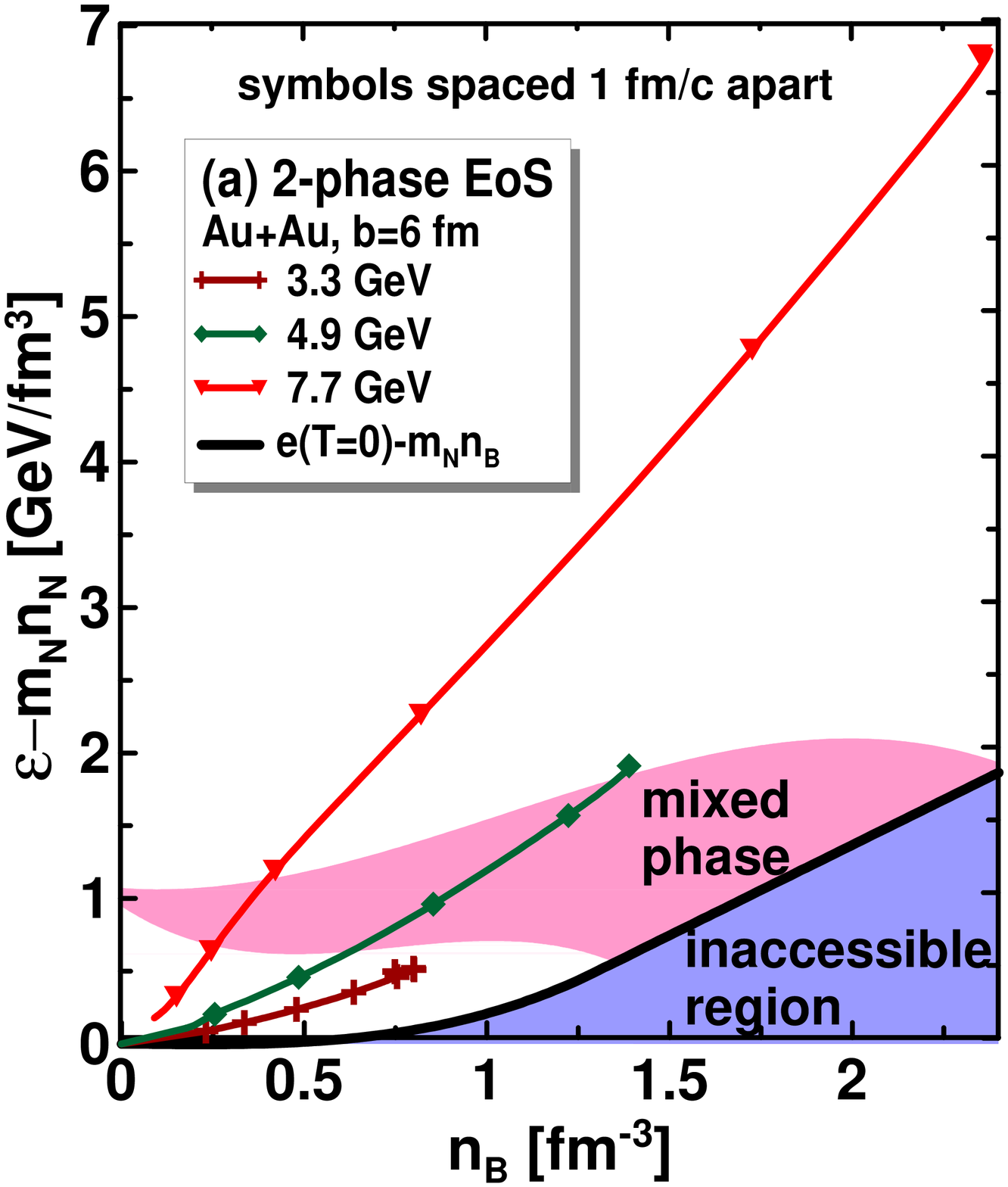}
\includegraphics[width=5.8cm]{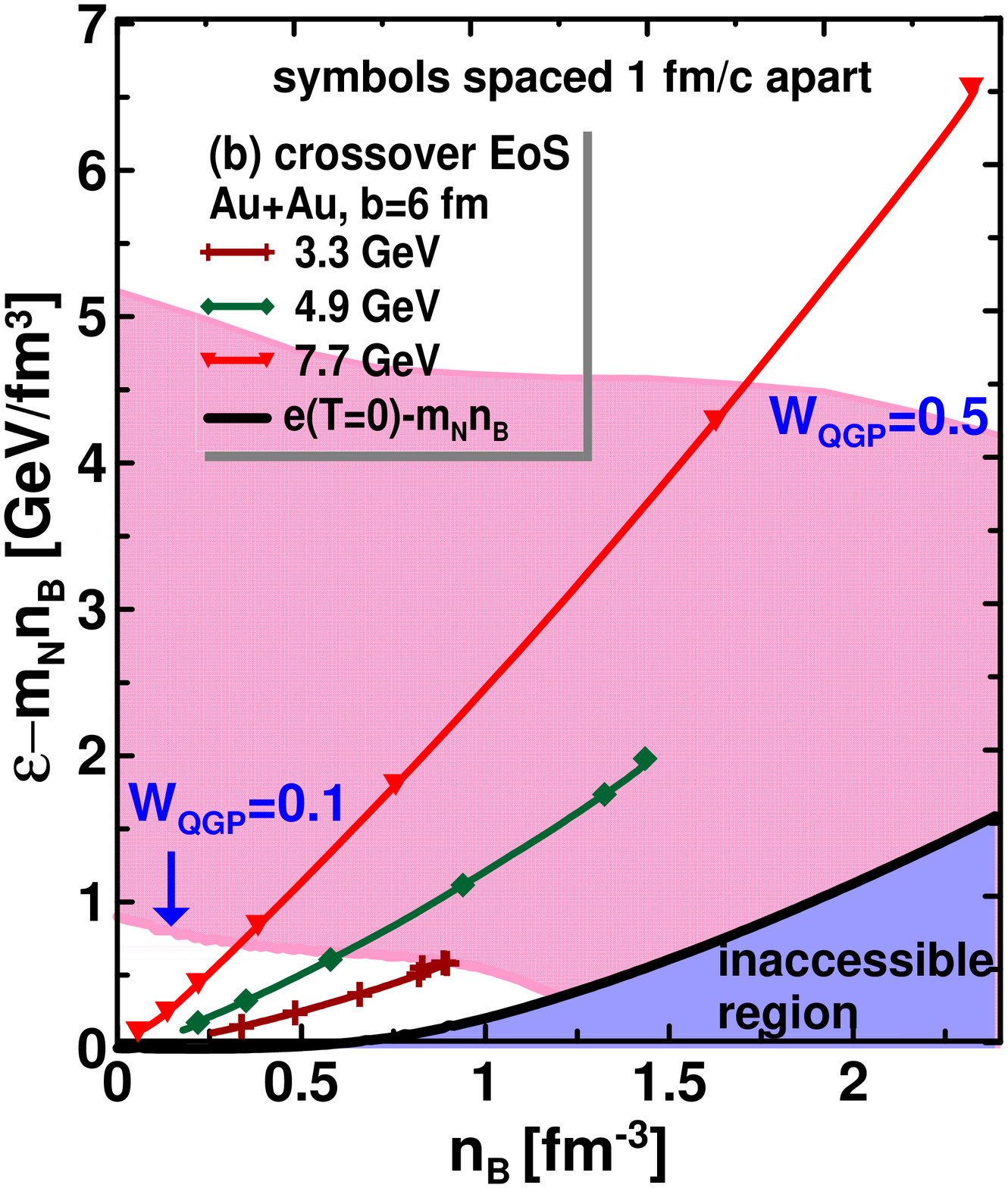}
 \caption{(Color online)
Dynamical trajectories of the matter in a central box of the
colliding nuclei  
(4fm$\times$4fm$\times \gamma_{cm}$4fm), where $\gamma_{cm}$ is the Lorentz
factor associated with the initial nuclear motion in the c.m. frame, 
for mid-central ($b=$ 6 fm) Au+Au collisions at 
$\sqrt{s_{NN}}=$ 3.3, 4.9 and 7.7 GeV. 
The trajectories are plotted in terms
of the baryon density ($n_B$) and 
the energy density minus $n_B$ multiplied by the nucleon mass 
($\varepsilon - m_N n_B$). 
Only expansion stages of the matter 
evolution are displayed.  
Symbols on the trajectories indicate the time rate of the evolution:
time span between marks is 1 fm/c. 
For the 2-phase EoS (a) 
the shadowed ``mixed phase'' region is located between the borders, 
where the QGP phase starts to raise ($W_{QGP}=$ 0) and becomes completely 
formed ($W_{QGP}=$ 1). 
For the crossover EoS  (b) the displayed borders correspond to
values of the QGP fraction $W_{QGP}=$ 0.1 and 0.5.
Inaccessible region is restricted by $\varepsilon(n_B,T=0)-m_N n_B$ from above. 
}
\label{fig1.0}
\end{figure}

Figure \ref{fig1.0} illustrates the EoS's used in the simulations and 
the onset of the deconfinement transition
in mid-central Au+Au collisions. 
Similarly to Ref. \cite{Randrup07}, 
the figure displays dynamical 
trajectories of the matter in the central box placed around the
origin ${\bf r}=(0,0,0)$ in the frame of equal velocities of
colliding nuclei. 
Initially, the colliding nuclei are placed symmetrically with respect
to the origin ${\bf r}=(0,0,0)$ along $z$ axis which is the direction of the beam.
At a given density $n_B$, the energy density $\varepsilon$ cannot be lower than 
the zero-temperature
compressional energy, $\varepsilon(n_B,T=0)$, 
so the accessible region is correspondingly limited. 
In the case of the crossover EoS only the region of the mixed phase between the QGP fraction 
$W_{QGP}=$ 0.1 and $W_{QGP}=$ 0.5 is displayed, because in fact the mixed phase
occupies the whole ($\varepsilon$-$n_B$) region. 
The $\varepsilon$-$n_B$ representation
is chosen because these quantities are suitable to compare
calculations with different EoS's. 
Only expansion stages of the dynamical trajectories are displayed, because the matter
in the box is close to thermalization then. 
The criterion of the thermalization%
\footnote{
strictly speaking, a randomization of the initial momentum 
because the matter is still chemically nonequilibrium
}
is the equality of longitudinal  and transverse
pressures in the box with an accuracy better than 10\%. 
These pressures are calculated by means of summation of the corresponding 
diagonal elements of the hydrodynamical energy-momentum 
tensor $T^{\alpha}_{ii}$ of ($i=p,t,f$)-fluids in the c.m. frame of
colliding nuclei
\begin{eqnarray*}
 \label{PzPtr}
P_{\scr{long}} = \sum_{\alpha=p,t,f} T^{\alpha}_{zz},
\\
P_{\scr{tr}} = \sum_{\alpha=p,t,f} (T^{\alpha}_{xx}+T^{\alpha}_{yy})/2.
\end{eqnarray*}
Evolution proceeds from the top point of the trajectory downwards.
Symbols mark the time intervals along the trajectory. 
Subtraction of the $m_N n_B$ term is taken for the sake of suitable 
representation of the plot. 
The size of the box was chosen 
to be large enough in order that the amount of matter in it can be
representative to conclude on the onset of deconfinement 
and to be small enough to consider the matter in it as a homogeneous
medium. Nevertheless, the matter in the box still amounts to a minor part
of the total matter of colliding nuclei.  
Therefore, only a minor part of the matter of colliding nuclei undergoes  the
deconfinement transition at 4.9 GeV energy within the 2-phase scenario. 
As seen, in the 2-phase scenario the deconfinement transition starts at the top AGS energies  
and gets practically completed at low SPS energies. 
In the crossover scenario it lasts till very high incident energies.

The 3FD model reproduces the major part of bulk observables, especially within deconfinement scenarios 
\cite{Ivanov:2013mxa,Ivanov:2013wha,Ivanov:2012bh,Ivanov:2013yqa,Ivanov:2013cba,Ivanov:2013yla}. 
This model does not include viscosity in its formulation. 
However, dissipation is present through the friction interaction between fluids. 
This dissipation is strongest at the early stage of the collision, when 
p- and t-fluids
 interpenetrate each other. At later stages, 
the baryon-rich 
(p- and t-)
fluids either get unified or spatially separated. Thus, their 
friction ceases to act. At the same time the net-baryon-free f-fluid survives as 
a separate instance till the very freeze-out. 
At high incident energies ($\sqrt{s_{NN}}\gsim$ 10 GeV), 
incomplete baryon stopping
results in spatial separation of the projectile-like and
target-like leading particles at late stages of the evolution, 
i.e. a transition from a single baryon-rich  
cluster of unified p- and t-fluids
at lower incident
energies to two spatially separated  
clusters consisting of spatially separated p- and t-fluids at higher energies,
takes place.  This was illustrated in Ref. \cite{Russkikh:2006aa}
for the hadronic scenario. For the deconfinement-transition scenarios the
picture is similar.
Therefore, at the freeze-out stage
the net-baryon-free f-fluid, predominantly located in the center region, 
is well separated in space from the baryon-rich 
(p- and t-) ones at
$\sqrt{s_{NN}}\gsim$ 10 GeV. It only overlaps 
with a small fraction of the unified baryon-rich fluid stopped near midrapidity, 
the amount of this unified stopped matter 
gradually decreases with the incident energy rise. Therefore, the 
dissipation produced by friction between unified baryon-rich fluid 
consisting of p- and t-ones,
and the net-baryon-free f-fluid  
at the late stage of the collision also 
gradually decreases with the energy rise.
In terms of an effective viscosity of the multi-fluid system it implies that 
the viscosity also gradually decreases.
Though it is highly difficult 
to quantitatively express the 3FD dissipation in terms of the effective viscosity, 
because this dissipation  depends on the dynamics of the collisions rather then only on 
the parameters of the 3FD model.

At lower incident energies ($\sqrt{s_{NN}}\lsim$ 10 GeV) the overlap between baryon-rich 
(p and t)
and net-baryon-free (f) 
fluids is stronger, however, the density of the f-fluid becomes 
lower. Therefore, it is not obvious that the dissipation is strong.

Any case, the elliptic flow is the most sensitive quantity to the dissipation effects. 
Comparison of the 3FD predictions with experimental data on the elliptic flow 
should indicate how relevant the above mechanism of dissipation is.

The elliptic flow 
is proportional to the spatial anisotropy \cite{Voloshin:2008dg,Ollitrault:1992bk} 
usually described by an eccentricity $\varepsilon$ 
defined as 
\begin{eqnarray}
 \label{eps}
\varepsilon = \frac{\langle y^2\rangle - \langle x^2\rangle}%
{\langle y^2\rangle + \langle x^2\rangle}\,\,,
\end{eqnarray}
where $\langle x^2\rangle$ and $\langle y^2\rangle$ are 
mean square values of spacial transverse coordinates in and
out of the reaction plane, respectively. 
These mean values are usually calculated
with either the wounded-nucleon~(WN) or the binary-collision~(BC) weights,
for details see Ref. \cite{Jacobs00}. 
These calculations are
based on the usual Woods--Saxon profile of the nuclear density  
\begin{eqnarray}
 \label{Woods-Saxon}
\rho (r) = \frac{\rho_0}{1+\exp[(r-R_A)/d]}, 
\end{eqnarray}
where $\rho_0$ is the normal nuclear density, $R_A=1.12 A^{1/3}$ is the radius of a nucleus with a mass number $A$, 
and $d$ is a diffuseness of the nuclear surface. 
As long as the eccentricity is small, elliptic flow should be directly proportional
to the eccentricity. For numerically
large eccentricities the direct proportionality could break in principle, but as was
shown in the very first hydrodynamic calculation by Ollitrault \cite{Ollitrault:1992bk} 
the proportionality holds well even for rather large values of $\varepsilon$.

\begin{figure}[thb]
\includegraphics[width=6cm]{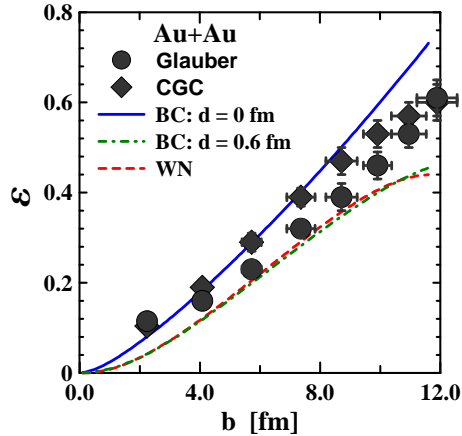}
\caption{(Color online)
Spatial eccentricity
$\varepsilon$ as a function of impact parameter in Au+Au collisions
for different surface diffusenesses ($d$) of the Au nucleus and
different weights of averaging: the wounded-nucleon (WN) and
the binary-collision (BC) weights~\cite{Jacobs00}. 
The results on   $\varepsilon_{\rm part}\{2\}$ [see Eqs.  
(\ref{eps-part}) and  (\ref{sigma-x-y})] 
are also displayed. Theses are calculated in Ref. \cite{Adamczyk:2012ku}
within the Monte-Carlo Glauber and Color Glass
Condensate (CGC) models. }
\label{fig1}
\end{figure}

Within the 3FD model the initial nuclei are represented by sharp-edged
spheres, i.e. with zero diffuseness ($d=0$). This is done for stability of 
the incident nuclei before collision. However, this approximation essentially affects 
the eccentricity. 
The  results obtained with $d=0$ and the realistic value of $d=$ 0.6 fm 
calculated with BC  weights
are shown in Fig.~\ref{fig1}. As seen, the $(d=0)$-result noticeably exceeds 
the eccentricity for the realistic value of $d=$ 0.6 fm. 
The ($d=0.6$ fm)-result with BN weights practically coincides with 
the eccentricity calculated with WN weights that 
is accepted as a default eccentricity in the 
experimental analysis \cite{Jacobs00}. 
The overestimation of $\varepsilon$ in the 3FD model naturally causes  
a respective overestimation of the elliptic flow. 
In order to resolve this problem, 
the calculated values of $v_2$ are rescaled 
with the factor of $\varepsilon_{BN}(d=0.6 \mbox{ fm})/\varepsilon_{BN}(d=0)$.  
This recipe does not imply that eccentricities calculated within the Glauber model
and the 3FD model necessarily coincide. It  only assumes that the eccentricities as a function
of the surface diffuseness change similarly within the Glauber 
and 3FD models, more precisely, 
$\varepsilon_{BN}(d=0.6 \mbox{ fm})/\varepsilon_{BN}(d=0) \approx 
\varepsilon_{3FD}(d=0.6 \mbox{ fm})/\varepsilon_{3FD}(d=0)$. 
This assumption is based on the fact that 
at a fixed impact parameter the eccentricity 
decreases with diffuseness rise due to simple geometrical reasons which hold 
true for both models.  
In the earlier works on the elliptic flow with the purely hadronic EoS
\cite{3FDflow,3FD-GSI07,3FDv2} this rescaling was not applied, because 
its need was realized only recently.

Figure~\ref{fig1} also displays the root-mean-square participant
eccentricity, $\varepsilon_{\rm part}\{2\}$, defined as \cite{Adamczyk:2012ku}
\begin{eqnarray}
 \label{eps-part}
\varepsilon_{\rm part} = \frac{\sqrt{(\sigma_y^2 - \sigma_x^2)^2+4\sigma_{xy}^2}}%
{\sigma_y^2 + \sigma_x^2}, \quad
\varepsilon_{\rm part}\{2\} = \sqrt{\langle\varepsilon_{\rm part}^2\rangle}
\end{eqnarray}
\begin{eqnarray}
 \label{sigma-x-y}
\sigma_x^2 &=& \{x^2\} - \{x\}^2,\quad \sigma_y^2 = \{y^2\} - \{y\}^2,  
\cr
\sigma_{xy} &=& \{xy\} - \{x\}\{y\},  
\end{eqnarray}
where the curly brackets denote the average over all participants
per event, while $\langle ...\rangle$, the average over events, and x and y are the positions of
participant nucleons.
Thus defined eccentricity takes into account event-by-event fluctuations 
with respect to the participant plane  
caused by a finite number of participant particles. 
The results displayed in Fig.~\ref{fig1} are calculated in Ref. \cite{Adamczyk:2012ku}
with the Monte-Carlo Glauber model \cite{Miller:2003kd,Miller:2007ri} and the Color Glass
Condensate (CGC) model \cite{Drescher:2006pi,Drescher:2006ca,Hirano:2009ah}. 
Correspondence between experimental centrality and impact parameter is taken from Ref.   
\cite{Adams:2005ca}
where the mean values of the impact parameter were obtained
using a Monte-Carlo Glauber calculation.

The displayed results on $\varepsilon_{\rm part}\{2\}$ demonstrate two points. 
First, there is an uncertainty related to the rescaling described above. 
Only for mid-central collisions ($b=$ 5$-$9), $\varepsilon_{BN}(d=0.6 \mbox{ fm})$ 
is close to  $\varepsilon_{\rm part}\{2\}$(Glauber). At smaller and bigger impact 
parameters the difference is substantial. 
Second, possible effects of the fluctuations 
resulting from a finite number of participant particles could be significant beyond
the range of mid-central collisions. 
Certainly, such kind of fluctuations are beyond the scope of the 3FD model 
because it deals only with continuous-medium quantities.

At RHIC and LHC energies 
the above mentioned Glauber and CGC models are conventionally used 
to prepare an initial state for the further hydrodynamic treatment of the 
expanding system. The advantage of thus constructed initial state is that it 
takes into account random fluctuations. Though, such an initial state is ``static'', 
i.e. the initial collective motion is disregarded. 
However, an advanced version of the Glauber model \cite{Vovchenko:2014gda} 
overcomes the latter shortcoming, it is able to take into account large accumulated 
angular momentum and shear flow that affect the directed flow and other odd
harmonics of the collective flow \cite{Csernai:2014cwa}.

\section{Integrated Elliptic Flow of Charged Particles}
\label{Charged}

Calculations of the integrated elliptic flow of charged particles were performed at fixed 
impact parameters $b$ which relate to experimental centralities as described above \cite{Adams:2005ca}. 
The integration over transverse momentum ($p_t$) was cut from above, $p_t< 2$ GeV/c. 
This constraint was chosen because the STAR data
were taken with this acceptance. Another reason is that the hydrodynamic treatment 
becomes inapplicable at high transverse momenta, as claimed in Ref.
\cite{Heinz:2009xj}, already at $p_t> 1.5$ GeV/c.

\begin{figure}[bt]
\includegraphics[width=8.50cm]{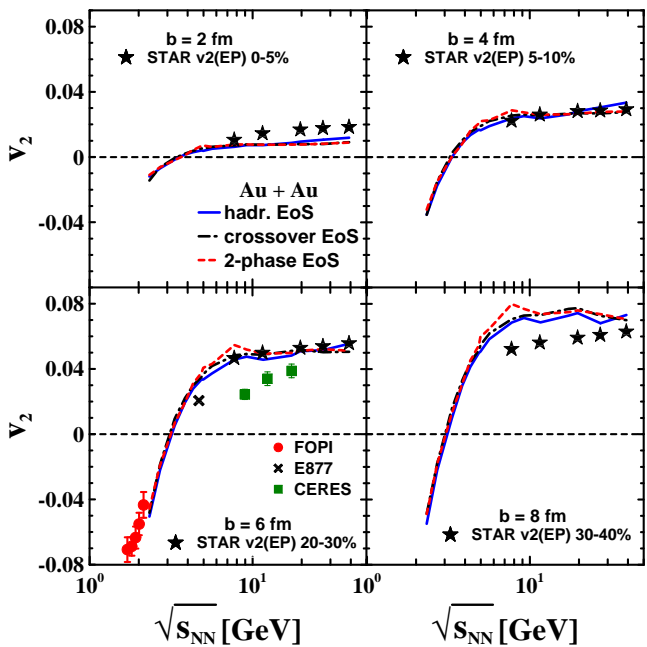}\\
 \caption{(Color online)
Elliptic flow of charged particles
at midrapidity as a function of incident energy
in collisions 
Au+Au at various centralities (impact parameters, $b$) obtained under the constraint of $p_t< 2$ GeV/c. 
Experimental data on the integral elliptic flow of charged particles are from  
STAR  Collaboration  \cite{Adamczyk:2012ku} (subset v2(EP)). 
FOPI data for Z=1 particles \cite{FOPI05}, as well as E877 \cite{E877} 
and CERES \cite{Adamova:2002qx} data for  charged particles  
are also displayed. 
} 
\label{fig2}
\end{figure}

\begin{figure}[bt]
\includegraphics[width=8.50cm]{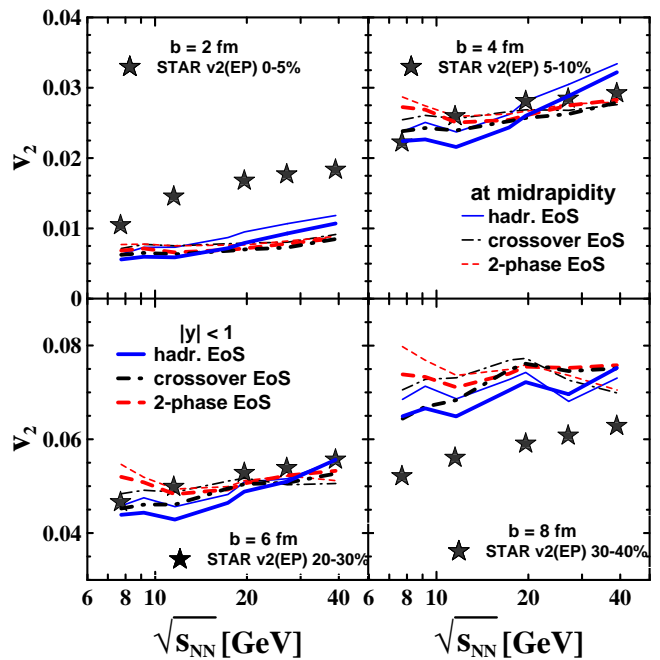}\\
 \caption{(Color online)
The same as in Fig. \ref{fig2} but with zoomed regions of the STAR 
data \cite{Adamczyk:2012ku}. In addition to the 
midrapidity $v_2$ (thin lines), the elliptic flow averaged over 
rapidity region $|y|<1$ is presented (bold lines). 
} 
\label{fig2a}
\end{figure}

Comparison of the integrated elliptic flow of charged particles
at midrapidity calculated within different scenarios with 
STAR  data \cite{Adamczyk:2012ku} 
for different centralities is presented in Fig. \ref{fig2}. 
Only the $v_2\{EP\}$ subset of data is presented because  
other data subsets are quite indistinguishable from the former one 
within the scale of this figure. 
As seen, all different scenarios give almost identical results which 
perfectly agree with the data for mid-central collisions 
(centralities 5-30\% or $b=$ 4$-$6 fm). 
However, in view of above discussed uncertainties of the 
eccentricity rescaling and  
the fact that  
the STAR data \cite{Adamczyk:2012ku} actually  correspond 
to the elliptic flow averaged over pseudorapidity region $|\eta|<1$%
\footnote{
see the discussion, concerning Fig. \ref{fig2a}, below
}
this agreement should be considered as simply good.

Again relying on discussion in the previous section, the observed 
underestimation of the elliptic flow at $b=$ 2 fm can be associated with
fluctuations resulting from a finite number of participant particles. 
The calculated $v_2$ is approximately  twice as 
low 
as compared with data. 
The subset v2(EP) of the STAR data is evaluated versus an event plane,
which generally does not coincide with the reaction plane, 
with respect to which the 3FD $v_2$ is determined. 
The difference results from fluctuations, the effect of which is
especially strong in near-central collisions. 
As seen from Fig.~\ref{fig1}, the eccentricity $\varepsilon_{\scr{WN}}$ at $b=$ 2 fm
calculated without fluctuations  is approximately twice as 
low 
as compared with
$\varepsilon_{\rm part}\{2\}$ that takes into account event-by-event
fluctuations. 
In particular, the fluctuations make the event-plane $v_2$ value non-zero 
even for head-on collisions, while the non-fluctuating reaction-plane $v_2$ 
is identically zero in this case. 
At more peripheral collisions
the fluctuation effect is weaker, as it also seen from Fig.~\ref{fig1}, and 
therefore the applicability of the 3FD (fluctuation-free) model is better. 
The observed overestimation of the elliptic flow at $b=$ 8 fm is expected. 
At large impact parameters, the number of particles in the participant zone 
becomes small. Therefore, the applicability of the hydrodynamics becomes worse. 

As for the predicted non-monotonicity of the integrated $v_2$ \cite{Kolb:2000sd} as a function 
of $\sqrt{s_{NN}}$, it indeed takes place for first-order-transition scenario: see 
a weak pick at $\sqrt{s_{NN}}\approx$ 8 GeV and the subsequent fall. 
Although, this non-monotonicity is very weak for charged particles.  
It is not observed in data.

\begin{figure*}[bht]
\includegraphics[width=13.70cm]{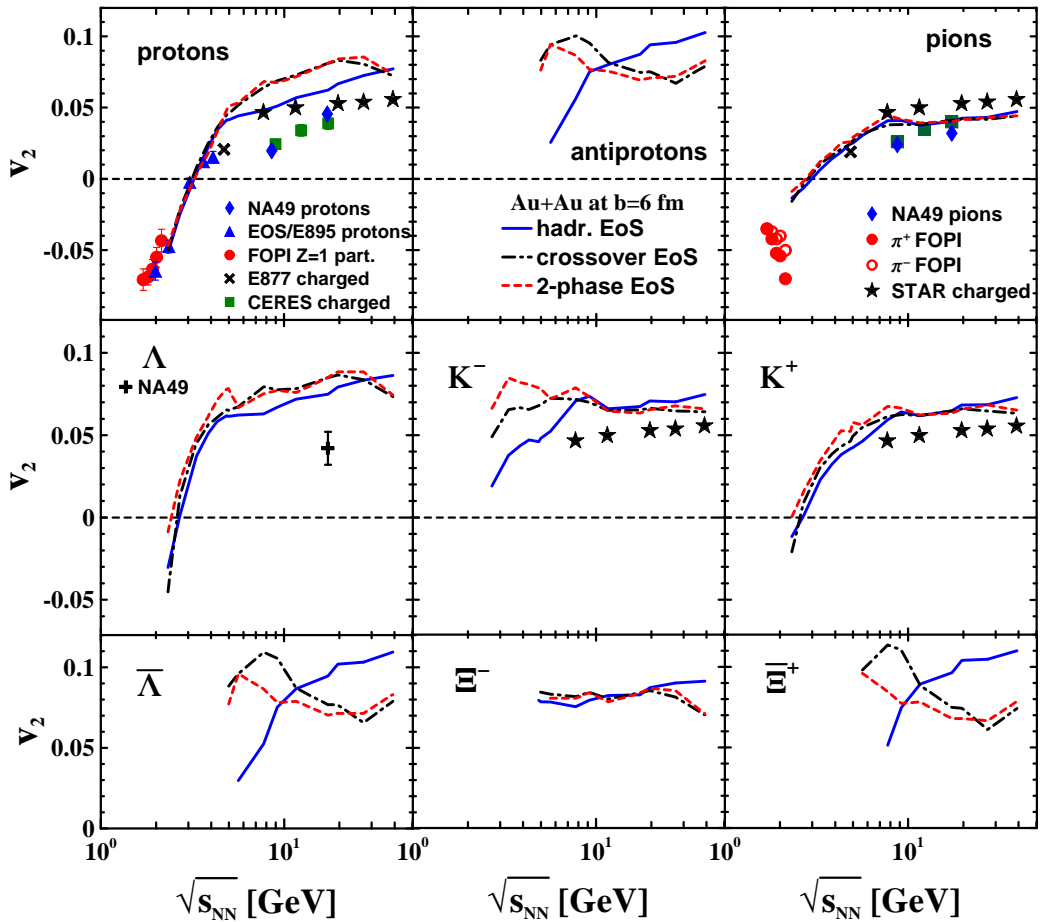}\\
 \caption{(Color online)
Transverse-momentum integrated 
elliptic flow of various hadronic species
at midrapidity as a function of incident energy
in mid-central collisions 
Au+Au at $b=6$ fm  obtained under constraint $p_t< 2$ GeV/c. 
Experimental data on the integral elliptic flow are from  FOPI Collaboration 
for Z=1 particles \cite{FOPI05}, EOS
and E895 \cite{Pinkenburg:1999ya} for protons, 
E877 \cite{E877} 
and CERES \cite{Adamova:2002qx} for all charged particles,   
NA49 for protons \cite{Alt:2003ab}, pions \cite{Alt:2003ab}, and $\Lambda$ hyperons 
\cite{Alt:2006ye}, and STAR \cite{Adamczyk:2012ku} (subset v2(EP) 
at 20-30\% centrality) for all charged particles. 
FOPI pion data are from \cite{FOPI07}. 
} 
\label{fig3}
\end{figure*}

In Fig. \ref{fig2},  
FOPI data for Z=1 particles \cite{FOPI05}, as well as E877 \cite{E877} 
and CERES \cite{Adamova:2002qx} data for charged particles  
are also displayed. The CERES data appreciably differ from those of the 
STAR collaboration. Recently there were published new CERES data \cite{Adamova:2012md}, 
however, only on differential $v_2$. 
The FOPI data are included because Z=1 particles dominate among charged particles
in the respective energy range. 
As seen from Fig. \ref{fig2},  all considered scenarios properly describe the 
change of sign of the elliptic flow at the incident energy decrease and 
approach the  FOPI data. 
It is even better seen in Fig. \ref{fig3}, where proton $v_2$ is presented 
with additional experimental data. 
These negative values are 
a consequence of the squeeze-out effect resulting from blocking of the expanding 
central  
blob, consisting of the unified p- and  t-fluids%
\footnote{
The   f-fluid is negligible at these collision energies.},
by the spectator matter.

Figure \ref{fig2a} displays zoomed regions of the STAR 
data \cite{Adamczyk:2012ku}. In addition to the 
midrapidity $v_2$ of charged particles (thin lines), 
the elliptic flow of charged particles averaged over a rapidity region $|y|<1$ 
is also presented (bold lines): 
\begin{eqnarray}
\label{|y|<1}
v_2 (|y|<1) = \int_{-1}^1 dy \, v_2 (y) \frac{dN}{dy} \slash \int_{-1}^1 dy \, \frac{dN}{dy}
\end{eqnarray}
where $dN/dy$ is a rapidity distribution of charged particles.  
This is done because the STAR data \cite{Adamczyk:2012ku} in fact correspond 
to the elliptic flow of charged particles averaged over a pseudorapidity region $|\eta|<1$.  
Of course, conditions $|y|<1$ and $|\eta|<1$ are not identical. However, the $|y|<1$
condition is a good approximation to $|\eta|<1$ at high incident energies, 
when charged particles are dominated by pions. 
As seen from Fig. \ref{fig2a}, the difference between midrapidity $v_2(y=0)$ and averaged $v_2 (|y|<1)$ 
values of the elliptic flow is not large. Moreover, this difference is comparable with the accuracy 
of computation ($\lsim$ 15\%). This justifies the comparison of 
the STAR data \cite{Adamczyk:2012ku} with the calculated midrapidity $v_2$ in Fig. \ref{fig2}.

\section{Integrated Elliptic Flow of Identified Hadrons}
\label{Integrated}

Let us turn to the elliptic flow integrated over transverse momentum 
for various species 
within different scenarios.
Results of such calculations are presented in Figs. \ref{fig3} 
for mid-central ($b=6$ fm) collisions of Au+Au. 
The integration over transverse momentum ($p_t$) is again cut from above, $p_t< 2$ GeV/c. 
Excitation functions of anti-baryons and heavy hyperons are displayed starting from 
energies above threshold of their production in the nucleon-nucleon collision. 
Below this threshold the hydrodynamic treatment of these species is inapplicable
because of low multiplicities of these species. 
The calculated $v_2$ values have accuracy not worse than 15\% 
(depending on the species and energy), as it was found in computations with finer grid. 
Results of the simulations presented in Fig. \ref{fig3}
are confronted to available data on the elliptic flow of protons, pions and lambdas. 
Data on the elliptic flow of charged particles are also displayed to guide an eye.

Here, predictions of alternative scenarios differ from each other to a different extent
depending on a particle. 
As seen from Fig. \ref{fig3}, 
the energy evolution of the midrapidity $v_2$ 
exhibits two basic patterns which also take place in excitation functions of 
the inverse slopes and mean transverse masses \cite{Ivanov:2013yla}. 
The first pattern (pattern I) is characteristic of baryons which populate all 
regions (both central and peripheral) of the excited system. In terms of the 3FD model, 
they originate predominately from the baryon-rich 
(p- and t-)  
fluids. Within this pattern, $v_2$ rises 
with energy sometimes beginning from negative values at low energies. 
As mentioned above, these negative values result from  
the squeeze-out effect.

The second pattern (pattern II) is characteristic of anti-baryons which are predominately produced 
in the central region of the excited system. In terms of the 3FD model, 
they originate  from the net-baryon-free f-fluid. The energy evolution of the anti-baryon $v_2$
is very distinct in purely hadronic and deconfinement scenarios. 
This distinction results from difference in dynamical evolution of the f-fluid 
that have already been discussed in Ref. \cite{Ivanov:2013yla} devoted to analysis of 
transverse-momentum spectra.

The dynamical evolution of the f-fluid is determined by three factors: a degree of stopping of 
colliding nuclei (i.e. the friction between them), the formation time ($\tau$) of the f-fluid, and the EoS 
itself. The  friction specifies the initial conditions, i.e. the energy deposit into the f-fluid. 
The formation time determines the beginning of the hydrodynamical expansion. Before it a collisionless 
expansion takes place. The EoS controls the character of the hydrodynamical expansion.
The first two quantities, the friction and $\tau$, were chosen on the condition of the best 
reproduction major part of bulk observables for each EoS, see Ref. \cite{Ivanov:2013wha,Ivanov:2013yqa}. 

The hadronic scenario is characterized by considerably longer formation time ($\tau=$ 2 fm/c) and 
stronger friction as compared to the deconfinement scenarios ($\tau=$ 0.17 fm/c). Therefore, 
in the hadronic scenario
the f-fluid exercises a longer collisionless expansion during which the spatial eccentricity of the 
system drops while the elliptic flow is not formed. The hydrodynamical expansion starts from 
essentially less deformed configuration as compared with that in the deconfinement scenarios. Thus, 
the hydrodynamically generated elliptic flow turns out to be lower than in the deconfinement scenarios 
at $\sqrt{s_{NN}}\lsim$ 10 GeV. At higher energies, two other factors (the friction and EoS)
come into game. The energy deposit into the f-fluid turns out to be higher  in the 
hadronic scenario than that for 
the deconfinement ones, which, in particular, is manifested in overestimation of anti-baryon 
and, somewhat later, meson production within the hadronic scenario \cite{Ivanov:2013yqa}. 
Besides, the deconfinement EoS's reduce the elliptic flow 
because they are softer than the hadronic one. 
As a result, the elliptic flow within the 
hadronic scenario becomes higher than that in the deconfinement ones.

Mesons demonstrate intermediate (between patterns I and II) behavior depending 
on whether they predominately produced in the central region or originate from 
both central and peripheral regions. Mesons, which require lower energy 
deposit for their production (pions and positive kaons), exhibit the $v_2$ 
behavior more similar to that of baryons (pattern I). At the same time, 
mesons, for production of which a higher  energy deposit is needed 
(negative kaons),  predominately originate from 
highly excited central region and hence their $v_2$ excitation functions 
are more similar to that of anti-baryons (pattern II). 

As already mentioned, the difference between predictions of the 
purely hadronic scenario and deconfinement ones is substantial 
for hadrons exhibiting pattern-II behavior. In particular, in this case 
the $v_2$ value indeed exhibits a non-monotonicity within the deconfinement
scenarios that was predicted in Ref. \cite{Kolb:2000sd}. 
The reason is the same as that discussed in Ref. \cite{Ivanov:2013mxa}
concerning the difference between the proton and antiproton  $v_2$. 
The proton $v_2$ at midrapidity is formed 
by particles from both the spatially center and   
peripheral regions of the nuclear system. 
This happens because the nuclear stopping is already quite strong 
at $\sqrt{s_{NN}}\leq$ 10 GeV, and hence the midrapidity quantities are determined  
not only by particles newly produced near the spacial center. 
The center and peripheral regions differently 
contribute to the midrapidity elliptic flow of different species, because 
they have different $v_2$ patterns. The interference between different $v_2$ patterns
washes out the non-monotonicity inherent in a separate pattern. 
At the same time, antiprotons are mostly produced 
from the central region with a definite $v_2$ pattern that survives  
in its midrapidity excitation function. 
This is also applicable to other antibarions and, to a lesser extent, to negative kaons.

However, the multiplicities of  anti-baryons 
are low at $\sqrt{s_{NN}}\leq$ 10 GeV. 
This results in large fluctuations of their $v_2$ which are, of course, beyond the scope 
of the 3FD model. This fluctuations
can reduce the observable $v_2$ and thus wash out the non-monotonicity. 
A possible destructive role of these fluctuations was indicated in Ref. \cite{Steinheimer:2012bn}.
It was shown that local fluctuations of the baryon number may lead to a biased 
determination of the event plane which may result in artificial reduction of 
antiproton $v_2$. The same mechanism of reduction is applicable to all other 
species of low multiplicity. 
The data on yet differential $v_2$ of antiprotons recently published
by STAR Collaboration \cite{Adamczyk:2013gw} apparently testify in 
favor of such scenario.

\section{Summary}
\label{Summary}

The integrated elliptic flow of charged particles from Au+Au collisions 
 was analyzed
in a wide range of incident energies
 2.7 GeV  $\le \sqrt{s_{NN}}\le$ 39 GeV. 
The analysis was done 
within the three-fluid model
\cite{3FD} employing three different EoS's: a purely hadronic EoS 
\cite{gasEOS}
and two versions of the EoS involving the deconfinement
 transition \cite{Toneev06}. These are an EoS with a first-order phase transition 
and that with a smooth crossover transition. 
It is found that all considered scenarios well reproduce 
recent STAR data \cite{Adamczyk:2012ku} on the integrated elliptic of charged particles
for mid-central Au+Au collisions. 
Moreover,  all considered scenarios properly describe the 
change of sign of the elliptic flow at the incident energy decrease below 
$\sqrt{s_{NN}}\approx$ 3.5 GeV. 
The problems met with central and peripheral collisions 
are naturally explained by restricted applicability of the 3FD model to those cases.

The present simulations  demonstrated that the integrated elliptic flow for charged particles at AGS-SPS-RHIC
energies 
reveals low sensitivity to
the EoS in agreement with 
the same observation made in Ref. \cite{Petersen:2010md,Dudek:2014qoa} for SPS and top RHIC energies.
Even within the first-order-transition scenario
the calculated elliptic flow  of charged particles practically does not exhibit
the non-monotonicity of the $v_2$ as a function 
of $\sqrt{s_{NN}}$ predicted in \cite{Kolb:2000sd}.  
This is a consequence of the nuclear stopping that is already substantial 
at $\sqrt{s_{NN}}<$ 10 GeV. Hence the midrapidity quantities are determined  
not only by particles newly produced near the spacial center. 
The center and peripheral regions differently 
contribute to the midrapidity elliptic flow because 
they have different $v_2$ patterns. The interference between different $v_2$ patterns
washes out the non-monotonicity inherent in a separate pattern.

The integrated elliptic flow of various species from Au+Au collisions  
was also predicted in simulations with the same three EoS's within the 
same energy range. A noticeable sensitivity to the EoS is found only for anti-baryons 
and, to a lesser extent, for 
$K^-$ mesons. 
In particular, in this case the $v_2$ excitation function 
indeed exhibits a non-monotonicity within the deconfinement
scenarios that was predicted in Ref. \cite{Kolb:2000sd}. 
Anti-baryons (and, to a lesser extent, 
$K^-$ mesons) are mostly produced 
in the central region with a definite $v_2$ pattern 
and their $v_2$ pattern is weakly affected by interference 
with those of peripheral regions. 
However, the multiplicities of  anti-baryons 
are low at $\sqrt{s_{NN}}\leq$ 10 GeV. 
This results in 
large fluctuations of their $v_2$. The fluctuations
can reduce the observable $v_2$ and thus wash out the non-monotonicity. 
A possible destructive role of these fluctuations was indicated in Ref. \cite{Steinheimer:2012bn}.
The data on yet differential $v_2$ of anti-protons recently published
by STAR Collaboration \cite{Adamczyk:2013gw} apparently testify in 
favor of such scenario.

\vspace*{3mm} {\bf Acknowledgements} \vspace*{2mm}

We are grateful to A.S. Khvorostukhin, V.V. Skokov,  and V.D. Toneev for providing 
us with the tabulated 2-phase and crossover EoS's. 
The calculations were performed at the computer cluster of GSI (Darmstadt). 
This work was partially supported  by  
grant NS-932.2014.2.


\begin{thebibliography}{999}
%
\bibitem{Adamczyk:2012ku} 
  L.~Adamczyk {\it et al.}  [STAR Collaboration],
  Phys.\ Rev.\ C {\bf 86}, 054908 (2012)
  [arXiv:1206.5528 [nucl-ex]].
%
\bibitem{Adamczyk:2013gw} 
  L.~Adamczyk {\it et al.}  [STAR Collaboration],
Phys. \ Rev.\ C {\bf 88} 14902 (2013)  
  [arXiv:1301.2348].
%
%
\bibitem{Kolb:2000sd} 
  P.~F.~Kolb, J.~Sollfrank and U.~W.~Heinz,
  Phys.\ Rev.\ C {\bf 62}, 054909 (2000)
  [hep-ph/0006129].
%
\bibitem{Kestin:2008bh} 
  G.~Kestin and U.~W.~Heinz,
  Eur.\ Phys.\ J.\ C {\bf 61}, 545 (2009)
  [arXiv:0806.4539 [nucl-th]].
%
\bibitem{Auvinen:2013sba} 
  J.~Auvinen and H.~Petersen,
  Phys.\ Rev.\ C {\bf 88}, 064908 (2013)
  [arXiv:1310.1764 [nucl-th]].
%
%
\bibitem{Petersen:2009vx} 
  H.~Petersen and M.~Bleicher,
  Phys.\ Rev.\ C {\bf 79}, 054904 (2009)
  [arXiv:0901.3821 [nucl-th]].
%
%
\bibitem{Petersen:2010md} 
  H.~Petersen and M.~Bleicher,
  Phys.\ Rev.\ C {\bf 81}, 044906 (2010)
  [arXiv:1002.1003 [nucl-th]].
%

\bibitem{Konchakovski:2012yg} 
  V.~P.~Konchakovski, E.~L.~Bratkovskaya, W.~Cassing, V.~D.~Toneev, S.~A.~Voloshin and V.~Voronyuk,
  Phys.\ Rev.\ C {\bf 85}, 044922 (2012)
  [arXiv:1201.3320 [nucl-th]].
%
%
\bibitem{Shen:2012vn} 
  C.~Shen and U.~Heinz,
  Phys.\ Rev.\ C {\bf 85}, 054902 (2012)
  [Erratum-ibid.\ C {\bf 86}, 049903 (2012)]
  [arXiv:1202.6620 [nucl-th]].
%
%
\bibitem{Solanki:2012ne} 
  D.~Solanki, P.~Sorensen, S.~Basu, R.~Raniwala and T.~K.~Nayak,
  Phys.\ Lett.\ B {\bf 720}, 352 (2013)
  [arXiv:1210.0512 [nucl-ex]].
%
\bibitem{Plumari:2013bga} 
  S.~Plumari, V.~Greco and L.~P.~Csernai,
  arXiv:1304.6566 [nucl-th].
%
%
\bibitem{Ivanov:2013mxa} 
  Yu. B.~Ivanov,
  Phys.\ Lett.\ B {\bf 723}, 475 (2013)
  [arXiv:1304.2307 [nucl-th]].
%
\bibitem{Steinheimer:2012bn} 
  J.~Steinheimer, V.~Koch and M.~Bleicher,
  Phys.\ Rev.\ C {\bf 86}, 044903 (2012)
  [arXiv:1207.2791 [nucl-th]].
%
%
\bibitem{Denicol:2013nua} 
  G.~S.~Denicol, C.~Gale, S.~Jeon and J.~Noronha,
  Phys.\ Rev.\ C {\bf 88}, no. 6, 064901 (2013)
  [arXiv:1308.1923 [nucl-th]].
%
%
\bibitem{Khvorostukhin:2010aj} 
  A.~S.~Khvorostukhin, V.~D.~Toneev and D.~N.~Voskresensky,
  Nucl.\ Phys.\ A {\bf 845}, 106 (2010)
  [arXiv:1003.3531 [nucl-th]].
%
\bibitem{Lacey:2013qua} 
  R.~A.~Lacey, A.~Taranenko, J.~Jia, D.~Reynolds, N.~N.~Ajitanand, J.~M.~Alexander, Y.~Gu and A.~Mwai,
  Phys.\ Rev.\ Lett.\  {\bf 112}, 082302 (2014)
  [arXiv:1305.3341 [nucl-ex]].
%
%
\bibitem{Dudek:2014qoa} 
  D.~M.~Dudek, W.~L.~Qian, C.~Wu, O.~Socolowski, S.~S.~Padula, G.~Krein, Y.~Hama and T.~Kodama,
  arXiv:1409.0278 [nucl-th].
%
%
\bibitem{3FD}
 Yu. B. Ivanov, V. N. Russkikh, and V.D. Toneev,
 Phys. Rev. C {\bf 73}, 044904 (2006) [nucl-th/0503088].
%
\bibitem{gasEOS}
V. M. Galitsky and I. N. Mishustin, Sov. J. Nucl. Phys. {\bf 29}, 181
(1979).
%
\bibitem{3FDflow}
V. N. Russkikh and Yu. B. Ivanov,
Phys. Rev. C {\bf 74} (2006) 034904
[nucl-th/0606007].
%
\bibitem{3FD-GSI07}
  Yu.~B.~Ivanov and V.~N.~Russkikh,
  PoS CPOD {\bf 07}, 008 (2007)
  [arXiv:0710.3708 [nucl-th]].
%
\bibitem{3FDv2}
       Yu. B. Ivanov, 
I. N. Mishustin, V. N. Russkikh, and L. M.~Satarov,
Phys. Rev. C {\bf 80}, 064904 (2009)
[arXiv:0907.4140 [nucl-th]].
%
%
%
%
%
%
\bibitem{Csernai:2011qq} 
  L.~P.~Csernai, D.~D.~Strottman and C.~Anderlik,
  Phys.\ Rev.\ C {\bf 85}, 054901 (2012)
  [arXiv:1112.4287 [nucl-th]].
%
%
\bibitem{Toneev06}
A. S. Khvorostukhin,  
V. V. Skokov, K. Redlich, and V. D. Toneev,
Eur. Phys. J. {\bf C48}, 531 (2006) [nucl-th/0605069].
%
\bibitem{Ivanov:2013wha} 
  Yu.~B.~Ivanov,
  Phys. Rev. C {\bf 87}, 064904 (2013) [arXiv:1302.5766 [nucl-th]]. 
%
\bibitem{Ivanov:2012bh} 
  Yu.~B.~Ivanov,
  Phys.\ Lett.\ B {\bf 721}, 123 (2013)
  [arXiv:1211.2579 [hep-ph]].
%
%
\bibitem{Ivanov:2013yqa} 
  Yu.~B.~Ivanov,
  Phys. Rev. C {\bf 87}, 064905 (2013) [arXiv:1304.1638 [nucl-th]]. 
%
\bibitem{Ivanov:2013cba} 
  Yu.~B.~Ivanov,
Phys. Lett. B {\bf 726}, 422  (2013) 
  [arXiv:1306.0994 [nucl-th]].
%
\bibitem{Ivanov:2013yla} 
  Yu. B.~Ivanov,
  Phys.\ Rev.\ C {\bf 89}, 024903 (2014)
  [arXiv:1311.0109 [nucl-th]]. 
%
%
\bibitem{Kat93}
U.~Katscher, D.H.~Rischke, J.A.~Maruhn, W.~Greiner,
I.N.~Mishustin, and L.M.~Satarov, Z. Phys. {\bf A346}, 209 (1993);
%
A.~Dumitru, U.~Katscher, J.A.~Maruhn, H.~St\"ocker, W.~Greiner,
and D.H.~Rischke, Phys. Rev.  C {\bf 51}, 2166 (1995) [hep-ph/9411358];
%
Z. Phys. {\bf A353}, 187 (1995) [hep-ph/9503347].
%
\bibitem{Brac97}
J.~Brachmann, A.~Dumitru, J.A.~Maruhn, H.~St\"ocker,
W.~Greiner, and D.H.~Rischke, Nucl. Phys. {\bf A619}, 391 (1997) [nucl-th/9703032];
%
M.~Reiter, A.~Dumitru, J.~Brachmann, J.A.~Maruhn, H.~St\"ocker,
and W.~Greiner, Nucl. Phys. {\bf A643}, 99 (1998) [nucl-th/9806010];
%
M.~Bleicher, M.~Reiter, A.~Dumitru, J.~Brachmann, C.~Spieles,
S.A.~Bass, H.~St\"ocker,  and W.~Greiner, Phys. Rev. C {\bf 59},
R1844 (1999)  [hep-ph/9811459].
%
%
%
\bibitem{Bass98}
S. Bass, {\em et al.},
Prog. Part. Nucl. Phys. {\bf 41}, 225 (1998) [nucl-th/9803035]. 
%
\bibitem{Cassing99}
  W.~Cassing and E.~L.~Bratkovskaya,
  Phys.\ Rept.\  {\bf 308}, 65 (1999).
%
\bibitem{CB09}
W. Cassing, E. L. Bratkovskaya, Nucl. Phys. A {\bf 831}, 215 (2009) [arXiv:0907.5331 [nucl-th]];
Phys. Rev. C {\bf 78}, 034919 (2008) [arXiv:0808.0022 [hep-ph]]; 
W. Cassing, Nucl. Phys. A {\bf 791}, 365 (2007) [arXiv:0704.1410 [nucl-th]].
%
\bibitem{Russkikh:2006aa} 
  V.~N.~Russkikh and Yu.~B.~Ivanov,
  Phys.\ Rev.\ C {\bf 76}, 054907 (2007)  [nucl-th/0611094].
%
\bibitem{Ivanov:2008zi} 
  Yu.~B.~Ivanov and V.~N.~Russkikh,
  Phys.\ Atom.\ Nucl.\  {\bf 72}, 1238 (2009)
  [arXiv:0810.2262 [nucl-th]].
%
\bibitem{Randrup07}
I. C. Arsene,  
L.V. Bravina, W. Cassing, Yu.B. Ivanov, A. Larionov, J. Randrup,
V.N. Russkikh, V.D. Toneev, G. Zeeb, D. Zschiesche, 
Phys. Rev. C {\bf 75}, 034902 (2007) [nucl-th/0609042].
%
%
\bibitem{Ollitrault:1992bk} 
  J.~-Y.~Ollitrault,
  Phys.\ Rev.\ D {\bf 46}, 229 (1992).
%
\bibitem{Voloshin:2008dg} 
  S.~A.~Voloshin, A.~M.~Poskanzer and R.~Snellings,
Elementary Particles, Nuclei and Atoms
(Springer-Verlag) {\bf 23}, 293 (2010)  [arXiv:0809.2949 [nucl-ex]].
%
\bibitem{Jacobs00}
P. Jacobs and G. Cooper, arXiv:nucl-ex/0008015.
%
\bibitem{Miller:2003kd} 
  M.~Miller and R.~Snellings,
  nucl-ex/0312008.
%
\bibitem{Miller:2007ri} 
  M.~L.~Miller, K.~Reygers, S.~J.~Sanders and P.~Steinberg,
  Ann.\ Rev.\ Nucl.\ Part.\ Sci.\  {\bf 57}, 205 (2007)
  [nucl-ex/0701025].
%
\bibitem{Drescher:2006pi} 
  H.~-J.~Drescher, A.~Dumitru, A.~Hayashigaki and Y.~Nara,
  Phys.\ Rev.\ C {\bf 74}, 044905 (2006)
  [nucl-th/0605012].
%
\bibitem{Drescher:2006ca} 
  H.~-J.~Drescher and Y.~Nara,
  Phys.\ Rev.\ C {\bf 75}, 034905 (2007)
  [nucl-th/0611017];
%
{\em ibid.} {\bf 76}, 041903 (2007)
  [arXiv:0707.0249 [nucl-th]].
%
\bibitem{Hirano:2009ah} 
  T.~Hirano and Y.~Nara,
  Phys.\ Rev.\ C {\bf 79}, 064904 (2009)
  [arXiv:0904.4080 [nucl-th]].
%
\bibitem{Vovchenko:2014gda} 
  V.~Vovchenko, D.~Anchishkin and L.~P.~Csernai,
  Phys.\ Rev.\ C {\bf 90}, no. 4, 044907 (2014)
  [arXiv:1407.4644 [nucl-th]].
%
\bibitem{Csernai:2014cwa} 
  L.~P.~Csernai and H.~Stoecker,
  J.\ Phys.\ G {\bf 41}, no. 12, 124001 (2014)
  arXiv:1406.1153 [nucl-th].
%
%
\bibitem{Adams:2005ca} 
  J.~Adams {\it et al.}  [STAR Collaboration],
  Phys.\ Rev.\ C {\bf 73}, 034903 (2006)
  [nucl-ex/0510053].
%
\bibitem{FOPI05}
A. Andronic {\it et al.} (FOPI Collaboration), Phys. Lett. {\bf B612},
173 (2005) 
[arXiv:nucl-ex/0411024].
%
\bibitem{E877}
P.~Braun-Munzinger and J.~Stachel,
  Nucl.\ Phys.\ A {\bf 638}, 3 (1998)
  [nucl-ex/9803015].
  %
\bibitem{Adamova:2002qx} 
  D.~Adamova {\it et al.}  [CERES Collaboration],
  Nucl.\ Phys.\ A {\bf 698}, 253 (2002).
%
\bibitem{Adamova:2012md} 
  D.~Adamova {\it et al.}  [CERES Collaboration],
  Nucl.\ Phys.\ A {\bf 894}, 41 (2012)
  [arXiv:1205.3692 [nucl-ex]].
%
\bibitem{Pinkenburg:1999ya} 
  C.~Pinkenburg {\it et al.}  [E895 Collaboration],
  Phys.\ Rev.\ Lett.\  {\bf 83}, 1295 (1999)
  [nucl-ex/9903010].
%
\bibitem{Alt:2003ab} 
  C.~Alt {\it et al.}  [NA49 Collaboration],
  Phys.\ Rev.\ C {\bf 68}, 034903 (2003)
  [nucl-ex/0303001].
%

\bibitem{Alt:2006ye} 
  C.~Alt {\it et al.}  [NA49 Collaboration],
  Phys.\ Rev.\ C {\bf 75}, 044901 (2007)
  [nucl-ex/0606026].
%
\bibitem{FOPI07}
W. Reisdorf, {\it et al.} (FOPI Collaboration),
Nucl. Phys. {\bf A781}, 459 (2007)   
[arXiv:nucl-ex/0610025]. 
%
%
\bibitem{Heinz:2009xj} 
  U.~W.~Heinz,
  in 'Relativistic Heavy Ion Physics', Landolt-Boernstein New Series, I/23, edited by R. Stock (Springer Verlag, New York,2010) Chap. 5
  [arXiv:0901.4355 [nucl-th]].

\end{thebibliography}
\end{document}